\documentclass[twocolumn,twocolappendix]{aastex63}
\usepackage{amsmath}
\usepackage{hyperref}
\usepackage{soul}

\newcommand\teff{$T_{\rm eff}$}
\newcommand\logg{$\log g$}

\newcommand\av{$A_V$}

\defcitealias{kounkel2018a}{Paper I}
\usepackage{listings}

\begin{document}
\title{ABYSS III: Observing accretion activity in young stars through empirical veiling measurements}

\author[0009-0004-9592-2311]{Serat Saad}
\affil{Department of Physics and Astronomy, Vanderbilt University, VU Station 1807, Nashville, TN 37235, USA}
\email{serat.mahmud.saad@vanderbilt.edu}

\author[0000-0002-5365-1267]{Marina Kounkel}
\affil{Department of Physics and Astronomy, University of North Florida, 1 UNF Dr, Jacksonville, FL, 32224, USA}
\email{marina.kounkel@unf.edu}

\author[0000-0002-3481-9052]{Keivan G.\ Stassun}
\affil{Department of Physics and Astronomy, Vanderbilt University, VU Station 1807, Nashville, TN 37235, USA}

\author[0000-0002-1379-4204]{A. Roman-Lopes}
\affil{Department of Astronomy, Universidad de La Serena, Av. Raul Bitran \#1302, La Serena, 170000, Chile}

\author[0000-0001-8600-4798]{C. Rom\'an-Z\'u\~niga}
\affil{Universidad Nacional Aut\'onoma de M\'exico, Instituto de Astronom\'ia, AP 106, Ensenada 22800, BC, M\'exico}

\author[0000-0001-6072-9344]{Jinyoung Serena Kim}
\affiliation{Steward Observatory, University of Arizona, 933 N. Cherry Ave., Tucson, AZ 85721-0065, USA}

\author[0000-0002-3389-9142]{Jonathan C. Tan}
\affil{Department of Astronomy, University of Virginia, Charlottesville, Virginia 22904, USA}
\affil{Dept. of Space, Earth \& Environment, Chalmers University of Technology, SE-41293 Gothenburg, Sweden}

\author[0000-0002-7795-0018]{R. Lo\'pez-Valdivia}
\affil{Universidad Nacional Aut\'onoma de M\'exico, Instituto de Astronom\'ia, AP 106, Ensenada 22800, BC, M\'exico}

\begin{abstract}
Stellar accretion plays an important role in the early stages of stellar evolution, particularly in Classical T Tauri Stars (CTTSs). Accretion of a CTTS can be related to different physical parameters such as effective temperature (\teff), age, abundance of hydrogen, etc. We can infer how accretion works by examining it across different wavelength regions. Accretion can be traced using veiling, a parameter that measures how excess emission from accretion affects the photospheric spectrum of CTTS. In this study, we selected a sample of CTTSs, Weak-line T Tauri Stars (WTTSs), and field stars, observed as a part of the SDSS-V Milky Way Mapper using the BOSS spectrograph. We measured veiling for CTTSs through comparing them to theoretical spectra. Next, we assessed the effect of veiling on different stellar properties, including wavelength, H$\alpha$ emission, effective temperature, and age. We investigated how veiling changes with these parameters and what the physical reasons behind the changes can be. Finally, we evaluated how our findings align with existing accretion shock models. This study highlights veiling as a critical diagnostic tool for understanding accretion in young stars.

\end{abstract}


\section{Introduction} \label{sec:intro}

Accretion plays a crucial role in the early stages of stellar evolution, particularly in a type of young stars known as Classical T Tauri Stars (CTTSs). These stars are magnetically active and accrete material from their surrounding circumstellar disks, which are the birthplaces of future exoplanets \citep{hartmann2016}. The timescale for planet formation coincides with that of star formation, and accretion processes significantly influence the lifespan of these disks, typically lasting a few million years \citep{hartmann2016}. Accretion not only affects the mass transfer to the star but also plays a crucial role in angular momentum evolution and jet launching \citep{fischer2023}. Therefore, understanding accretion is vital for developing comprehensive theories of stellar and planetary formation.

In the magnetospheric accretion model \citep{shu1994, romanova2002, bessolaz2008, hartmann2016}, the magnetic field of a CTTS truncates the inner disk at a few stellar radii, channeling the disk material into accretion columns that fall onto the star at nearly free-fall velocities. This process generates accretion shocks, leading to emission lines and excess continuum emission superimposed on the star's photospheric spectrum, a phenomenon known as veiling \citep{joy1949,calvet1998}. Veiling ($r_{\lambda}$) can be defined as following:

\begin{equation} \label{eqn:veiling}
    F_{\text{source, }\lambda} = \frac{F_{\text{phot, } \lambda} + r_{\lambda}}  {(1 + r_{\lambda})} 
\end{equation}

\noindent where, $F_{\text{source, }\lambda}$ is the total normalized flux coming from the source, $F_{\text{phot, } \lambda}$ is the normalized model flux, and $r_{\lambda}$ is the veiling \citep{basri1990hamilton, Kidder2021}.

High veiling values indicate substantial ongoing accretion, which can significantly influence the thermal and chemical properties of the disk and affect the conditions under which planets form \citep{morbidelly2016}.

The mass accretion rate ($\dot{M}_{\text{acc}}$) is a critical parameter in studying the evolution of pre-main sequence stars and their disks. It can be estimated from the accretion luminosity ($L_{\text{acc}}$), which is derived from accretion-powered emission lines using empirical relations \citep{gullbring1998, alcala2017}. Alternatively, $\dot{M}_{\text{acc}}$ can be inferred by comparing veiling values to accretion shock models. These models, initially one-dimensional and plane-parallel \citep{calvet1998}, have evolved to incorporate multiple energy fluxes and filling factors, allowing for complex accretion column structures \citep{ingleby2013, robinson2019, espaillat2022, pittman2022}. Optical veiling can also be used to infer accretion luminosity, as demonstrated by studies like \cite{stock2022}. Analyzing the relationship between veiling and stellar rotation provides further insights into the dynamics of accretion processes.

Traditionally, veiling has been assumed to be constant across different wavelengths, typically modeled using a handful of discrete lines \citep{Bertout1989, Gahm2008}. Only a few of the studies have been done where veiling has been compared with wavelength \citep{basri1990hamilton, ingleby2013}. Theoretical models suggest that veiling should vary as a function of wavelength due to the different physical conditions in the accretion stream and the circumstellar environment\citep{calvet1998}.

Empirical profiles of veiling across a broad range of wavelengths are sparse, limiting our understanding of accretion processes in young stars. So, assuming veiling to be constant across different wavelengths may not capture the true complexity of the accretion processes. Instead, understanding how veiling varies with wavelength can provide insights into the structure and dynamics of the accretion flows in CTTSs.

The Sloan Digital Sky Survey in its fifth iteration (SDSS-V, Kollmeier, J.A., et al. 2025, in prep), has obtained an unprecedented sample of optical and near IR spectra of millions of stars using the BOSS (Baryonic Oscillation Spectroscopic Survey) and APOGEE (Apache Point Observatory Galactic Evolution Experiment) spectrographs. The SDSS-V Milky Way Mapper (MWM) program, through the APOGEE \& BOSS Young Star Survey (ABYSS), aims to achieve high-resolution, multi-epoch spectra of over 100,000 young stars with ages less than 30 Myr. This provides a unique opportunity for statistical analysis of accretion processes \citep{kounkel2023abyss}. In the ABYSS program, more than 18,000 young stars, including over 5,500 CTTSs, have been identified using a newly developed method \citep{saad2024abyss}.

In this paper, we aim to use SDSS optical spectra to examine the veiling in CTTSs, its dependence on wavelength, and its variation with different physical properties such as effective temperature, age, and the equivalent width (eqw) of the H$\alpha$ line. In Section \hyperref[sec:data]{\ref{sec:data}}, we describe the data used for our analysis, including the sources of the spectra and the methods for extracting stellar parameters. In Section \hyperref[sec:analysis]{\ref{sec:analysis}}, we detail the analysis methods, including the correction for extinction and the calculation of veiling across different wavelength bins. We also discuss the empirical methods used to measure veiling without assuming a specific model. In Section  \hyperref[sec:results]{\ref{sec:results}}, we present the results of our veiling measurements, comparing veiling values across different temperature and age bins, and examining their consistency. Finally, in Section \hyperref[sec:discussion]{\ref{sec:discussion}}, we discuss the implications of our findings for understanding accretion processes in CTTSs and compare our results with theoretical models and previous empirical studies. We conclude in Section \hyperref[sec:conclusions]{ \ref{sec:conclusions}} with a summary of our findings and their significance for future research on accretion in young stellar objects.

\section{Data} \label{sec:data}

BOSS is a low-resolution spectrograph operated by the SDSS-V. There are two 2.5-meter telescopes: one located in the northern hemisphere at the Apache Point Observatory \citep[APO,][]{gunn2006} and the other in the southern hemisphere at the Las Campanas Observatory \citep[LCO,][]{bowen1973}. The strategic positioning of these observatories allows for comprehensive sky coverage, capturing spectra from both hemispheres. BOSS covers a wavelength range of 3600-10400 \AA, with a resolving power of R$\sim$1800 and a pixel scale of $\sim$1 \AA \citep{smee2013multi}. Each spectrograph at APO and LCO is capable of observing 500 spectra simultaneously within a field of view of 3$^\circ$ and 2$^\circ$, respectively. BOSS has a typical flux calibration precision of $\sim$10\% relative to the standard stars observed in the same field. This scatter is mostly driven by fiber positioning errors and variations in standard star distributions to be within 0.1. This level of precision is sufficient for reliable veiling measurements. The APOGEE and BOSS Young Star Survey (ABYSS) program within SDSS-V focuses on obtaining spectra of approximately 100,000 young stars \citep{kounkel2023abyss}. Since its initiation in 2021, BOSS has observed over a million stars, among which 40,000 of these targets being young stars identified by the ABYSS program.

We extracted fundamental stellar parameters of the sources observed by BOSS using BOSS Net, an advanced neural network. These parameters include effective temperature (\teff), surface gravity (\logg), metallicity ([Fe/H]), and radial velocity (RV) \citep{sizemore2024self}. The BOSS Net can predict \teff\ with a precision of 0.008 dex and \logg\ with a precision of 0.1 dex at a signal-to-noise ratio (SNR) of approximately 15, calibrated to theoretical models and specifically calibrated for YSOs. Accurate RVs are essential for precisely identifying spectral lines, especially those that are weak and narrow.

\begin{figure}
\epsscale{1.1}
\plotone{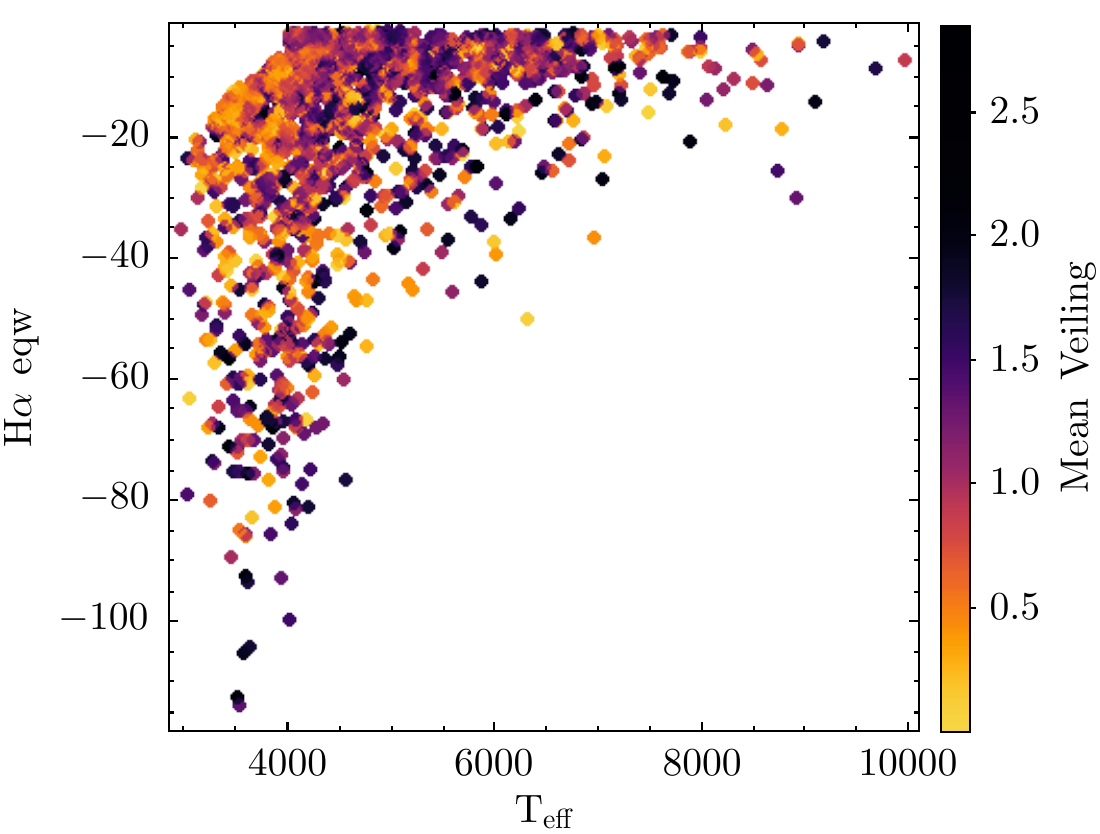}
\caption{Plotted H$\alpha$ eqw vs \teff\ of our sample color coded by mean veiling for each of the sources.
\label{fig:eqw_teff}}
\end{figure}

In \cite{saad2024abyss} we developed LINEFOREST, a pipeline to measure the equivalent widths and identifying the young stars based on them. Spectral lines, such as Li I, optical H I, and Ca II, are excellent indicators of stellar youth and can be used to distinguish young stars from other sources. These lines either appear in emission or absorption, depending on the physical conditions of the star and its surrounding environment. For example, the presence of the Li I 6708 \AA\ absorption line is a direct indicator of stellar youth, as lithium is rapidly depleted in stars older than a few million years \citep{jeffries2017}. Similarly, optical H I lines, such as H$\alpha$, often appear in strong emission in young stars due to accretion processes, and Ca II lines are associated with chromospheric activity, which is more prominent in younger stars \citep{briceno2019}. LINEFOREST can estimate the equivalent widths of 50 different youth-sensitive line regions of any optical spectra of a star. The pipeline then applies a set of criteria based on the properties of the spectral lines to identify young stars that have age $<$ 30 Myr. In \cite{saad2024abyss}, we also developed a technique that can predict whether any of these young stars are classical T Tauri stars (CTTSs) or not. This developed technique is based on the equivalent width of the H$\alpha$ line, a good indicator of strong accretion activity. Based on this pipeline and method we identified around 40,000 young stars and around 1900 CTTSs. Figure \ref{fig:eqw_teff} shows us the distribution of the \teff\, and H$\alpha$ eqw of the sources. We used these sources in this work to do further analysis to understand their accretion process.

We inferred the ages for each of these selected sources using Sagitta \citep{McBride2021sagitta}. Sagitta is a neural net that first identifies stars that are in pre-main sequence and then calculate their ages based on photometry, parallax, and average extinction.  Comparing to the traditional techniques that relies on theoretical isochrones for predicting ages, Sagitta offers more stable performance with
respect to \teff\ \citep{kounkel2023abyss}.

\section{Analysis} \label{sec:analysis}

To measure extinction and veiling in the spectra it is necessary to compare the spectrum of CTTS with another stellar spectrum that is not affected by either extinction or the accretion stream. 

There are two possible methods in extracting veiling - either 1) comparing the continuum level of the model and the observations, or 2) to look at the relative line depths between the model and the spectra. In the former case, it is common to use spectra neighboring WTTSs, however as the signal-to-noise of most observations in our sample is $<$30, doing so would have introduced significant noise in our calculation of veiling. It is also possible to use the synthetic spectra, however, with them, the first method would not be optimal due to various factors needed to properly calibrate the continuum, including calculating the radius of the star. As such, with synthetic spectra it becomes more efficient to work with fully normalized spectra and examine the effect of veiling in the line strengths instead.

Thus, we used synthetic PHOENIX spectra, which are high-resolution spectral templates created for a wide range of stellar parameters across various wavelength ranges (optical, near-infrared, infrared, and ultraviolet) \citep{husser2013new}. The PHOENIX library covers a parameter space adequate of all stages of stellar evolution: effective temperature (\teff) from 2300 K to 12,000 K, surface gravity (\logg) from 0.0 to +6.0, metallicity ([Fe/H]) from -4.0 to +1.0, and alpha enhancement ([$\alpha$/Fe]) from -0.2 to +1.2. We used [Fe/H] = 0 and [$\alpha$/Fe] = 0 for our models to ensure a consistent baseline, as these values represent solar metallicity and no alpha enhancement, which are typical for the young stars we are studying \citep{kounkel2023abyss, saad2024abyss}. Using these templates, we can perform spectral corrections and analyses on the observed CTTSs spectra.

\subsection{De-reddening Spectra}

\begin{figure*}
\epsscale{1.1}
\plotone{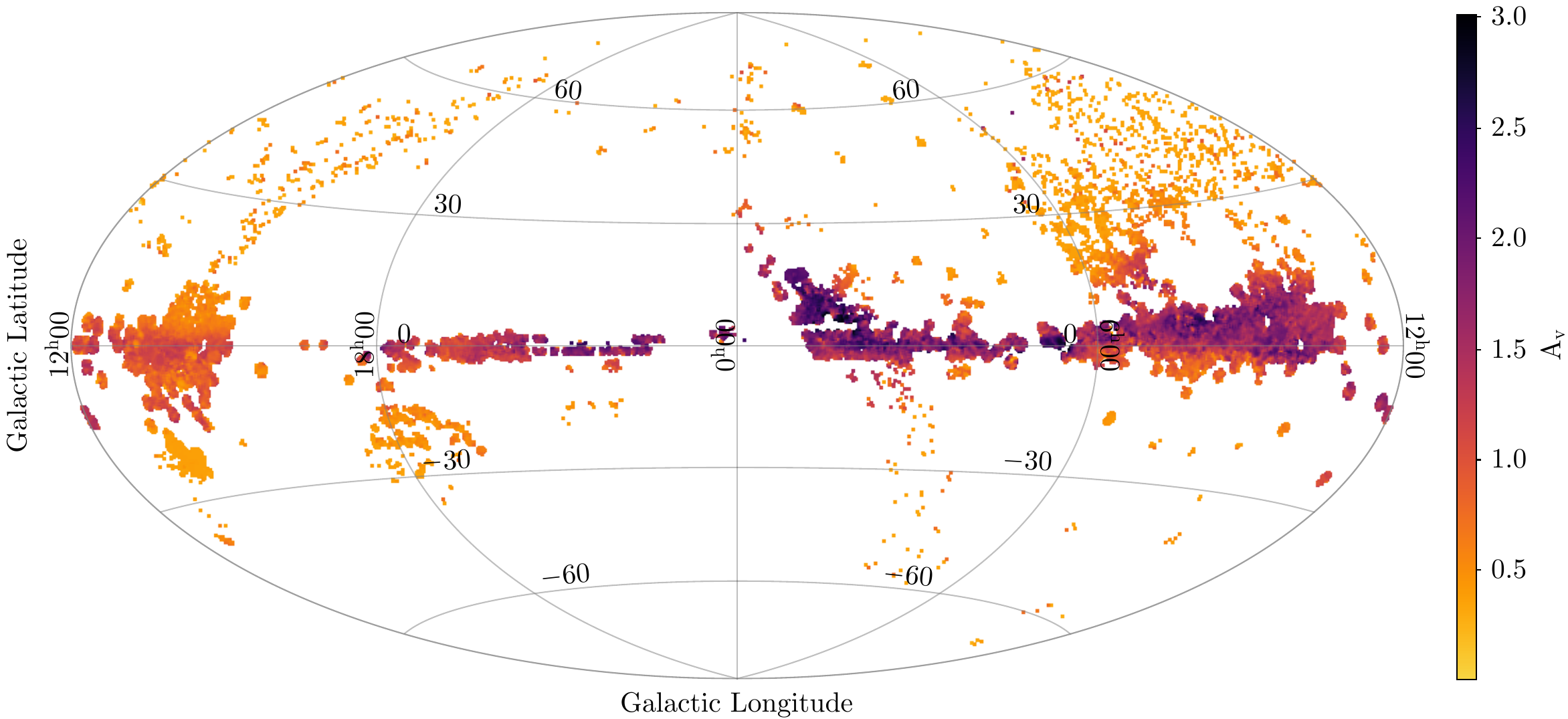}
\caption{On-sky distribution of identified YSO candidates with the calculated extinction parameter ($A_{V}$).
\label{fig:sky}}
\end{figure*}

\begin{figure}
\epsscale{1.1}
\plotone{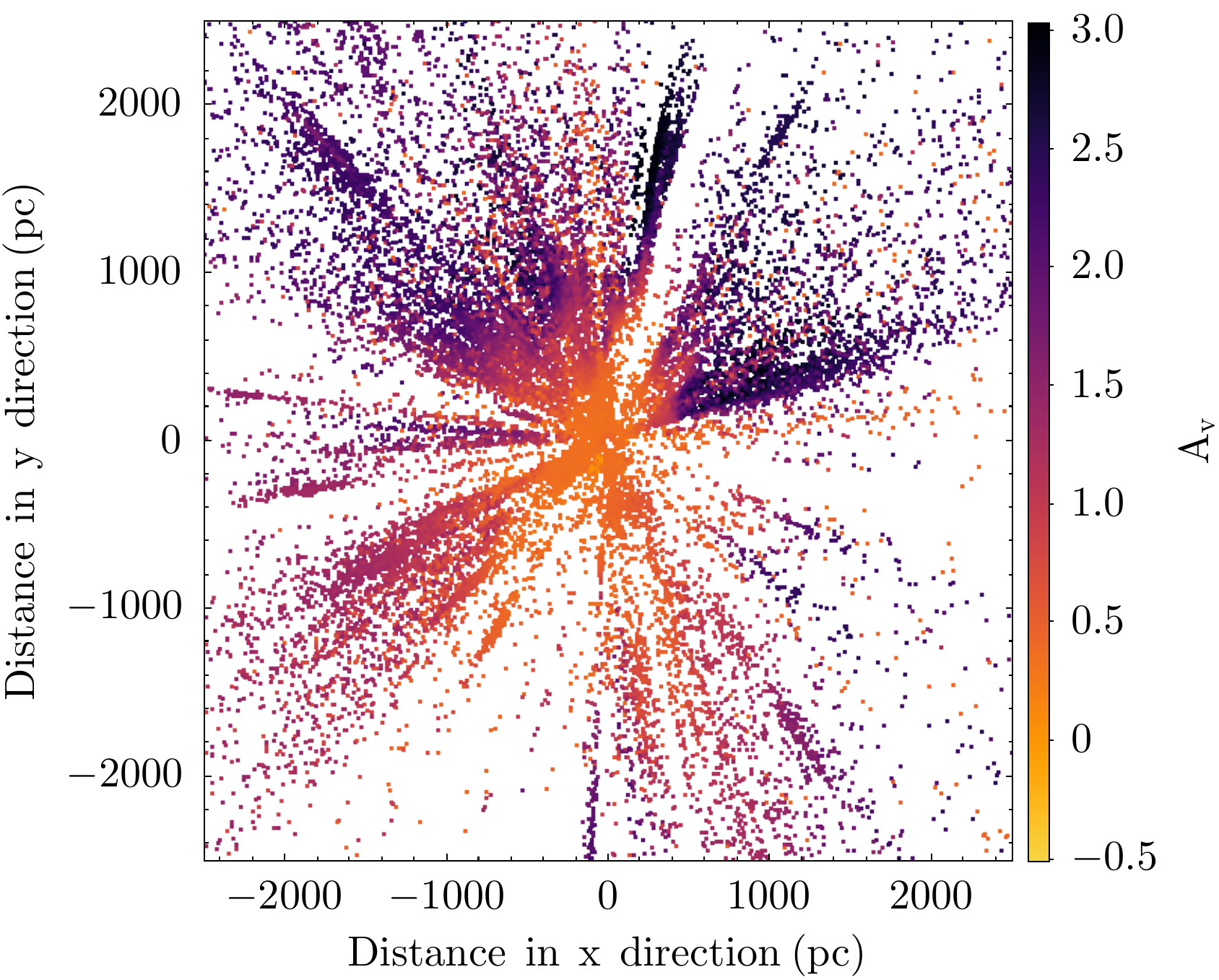}
\caption{Distribution of the sources in space color coded by the extinction parameter ($A_{v}$). 
\label{fig:dist_av}}
\end{figure}

The first step was to correct for extinction (reddening) in the spectra. Extinction is not the only one, but a dominant source of discrepancy between the synthetic model and the observed spectra other than the accretion. Though extinction can be stronger in blue and weaker in red, it can be characterized using a single coefficient (\av). The observed spectra were first matched with the corresponding synthetic spectra from the PHOENIX library. We first performed a sub grid interpolation between the observed spectra and the synthetic spectra to match physical parameters like \teff\ and \logg. Then we compared each observed spectrum to the corresponding synthetic spectrum using G23 model of python 'dust\_extinction' library \citep{dust_extinction}. This helped us to calculate the extinction parameter (\av) for each spectrum. Once the \av was determined, we de-reddened the spectra through dividing by the extinction profile. These de-reddened spectra are the ones we used for calculating veiling.

Figure \ref{fig:sky} shows the on sky distribution of identified young star candidates, with colors indicating the extinction parameter (\av). Higher extinction is concentrated along the Galactic plane where dust density is greater. Figure \ref{fig:dist_av} displays the spatial distribution of young star candidates in Cartesian coordinates also color coded by (\av). The radial pattern shows how extinction varies with distance.

\subsection{Measuring Veiling}

\begin{figure*}
\epsscale{1.1}
\plotone{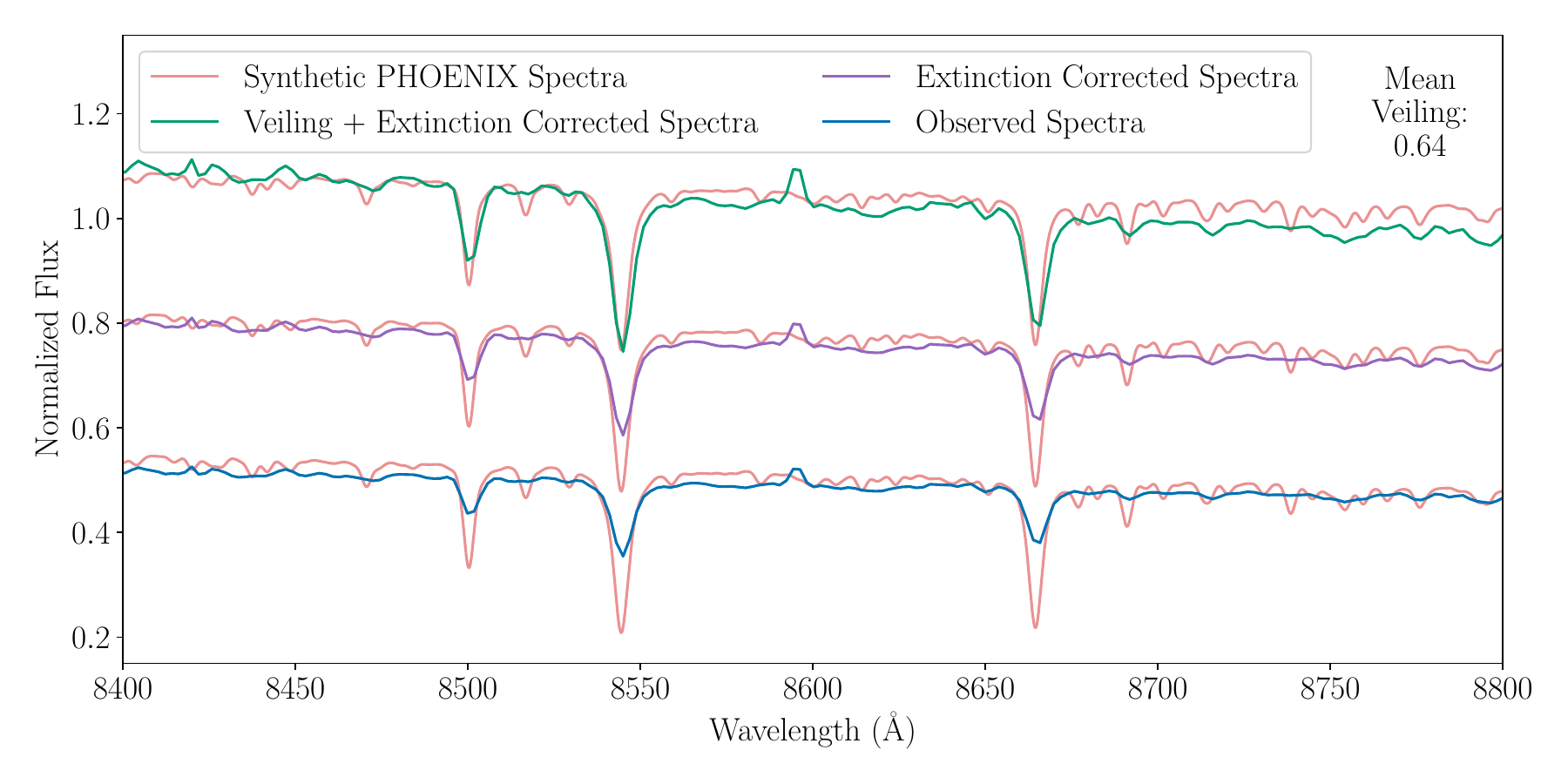}
\caption{Spectral correction of one of the sources. The observed spectra (in blue), extinction corrected spectra (in purple), and veiling + extinction corrected spectra (in green) has been plotted against the synthetic spectra model (in red). The veiling + extinction corrected spectra represent the photospheric spectra of the star. While veiling increases the photospheric flux, extinction decreases it. All the spectra are normalized and added to a constant for better visualization. The \teff\ of the source is 5694 K.
\label{fig:spectra_correction}}
\end{figure*}

\begin{figure}
\centerline{
    \includegraphics[width=\columnwidth]{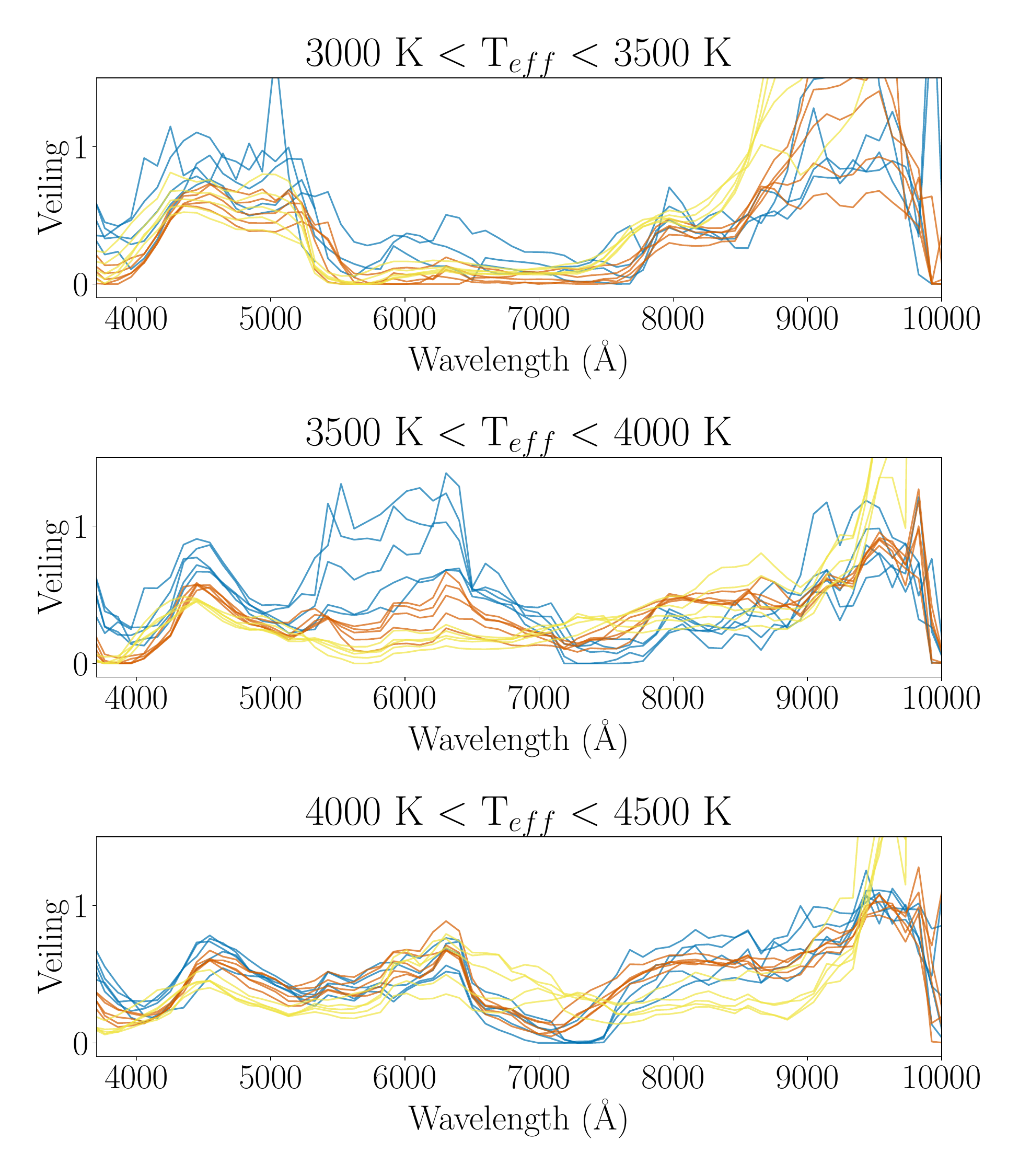} 
}
\caption{Veiling of CTTSs (in blue) and WTTSs (in red) and Non YSOs (in yellow) as a function of wavelength for different \teff\ ranges. Each of the lines represent a bin width of 100K \teff . We see WTTSs and Non YSOSs veiling due to spots and other factors of synthetic spectra.
\label{fig:veiling}}
\end{figure}

The next step is to measure the veiling profile for all the sources. Veiling ($r_{\lambda}$) can be defined using Equation \ref{eqn:veiling}. Veiling depends on wavelengths and, unlike \av, it cannot be characterized with a single constant. So, we empirically calculated veiling at various wavelength ranges without assuming any model for it. At first we interpolated the extinction corrected spectra with the corresponding synthetic spectra for each of the sources for a common wavelength grid. Then we divided both the synthetic and observed spectra into different wavelength bins. The wavelength bins were 1000 \AA\ wide with advancing in steps of 100 \AA, for the total BOSS spectra wavelength range of 3600-10000 \AA . Using individual lines to calculate veiling can limit us to isolated regions of the spectrum with missing subtle variations and wavelength dependent trends in veiling that could provide insights into the broader accretion structure. So instead of using individual lines to measure veiling, we chose an empirical approach across broader wavelength bins.

To calculate the veiling for each wavelength bin, we performed a least square optimization between the observed spectra and the corresponding PHOENIX template. In that case, each of the wavelength bins were allowed to have the best $r_{\lambda}$ for equation \ref{eqn:veiling}, where $F_{\text{phot, } \lambda}$ is the synthetic flux coming from the PHOENIX template. Figure \ref{fig:spectra_correction} illustrates the observed spectra (blue), extinction-corrected spectra (purple), and veiling + extinction-corrected spectra (green) compared to the synthetic PHOENIX spectra (red). The veiling + extinction-corrected spectra represent the star's photosphere, with veiling increasing and extinction decreasing the flux. All spectra are normalized for clarity.



We performed our analysis and veiling measurements for all the sources, including all of Classical T-Tauri Stars (CTTSs), Weak-lined T-Tauri Stars (WTTSs), and some field stars. Analysis for WTTSs and field stars may seem unnecessary because of the lack of accretion disk, but we performed the analysis for these sources to understand the systematic biases behind our approach.

In doing this we have obtained independent measurement of veiling ($r_{\lambda}$) for a total of $\sim$70 unique measurements per spectrum at different mean wavelengths for around 1800+ CTTSs, 35,000+ WTTSs, and selected 15,000+ field stars. Figure \ref{fig:veiling} displays the veiling profiles of Classical T-Tauri Stars (CTTSs), Weak-lined T-Tauri Stars (WTTSs), and non-YSOs across different wavelengths and effective temperature (\teff) ranges. The veiling is shown as a function of wavelength in bins of 100K for \teff. WTTSs and non-YSOs show veiling due to factors such as stellar spots. We also recorded the veiling uncertainty for each of the sources while calculating veiling.

\subsection{Correcting the veiling measurements}

\begin{figure}
\epsscale{1.1}
\plotone{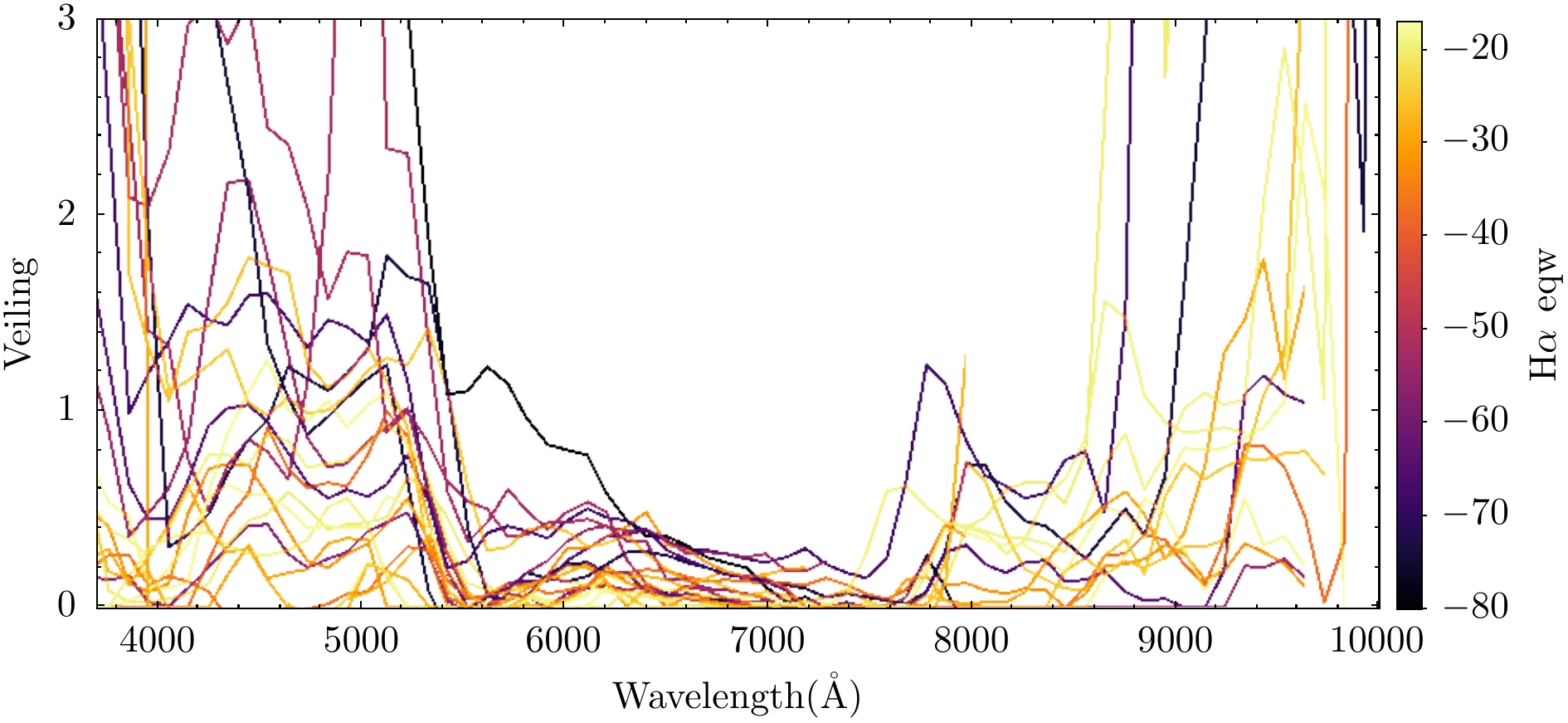}

\caption{Corrected Veiling vs Wavelength plot for a total 45 CTTSs among 1800+ sample which are in a very narrow (3330 K to 3450 K) \teff\ range color coded by the H$\alpha$ eqw before binning them. Each line here depicts one CTTS source. All these veiling values have been corrected by subtracting WTTSs veiling for the corresponding \teff\ range.
\label{fig:corrected_veiling}}
\end{figure}

The measured veiling can be affected by different systematics, some of which are wavelength dependent and some of which are dependent on the physical properties of stars like the \teff .

One such case is when the spectra are affected by infrared (IR) excess. IR excess in the stellar spectra can occur due to a dusty disk around the star. This excess is not accounted for in template matching and occurs due to a different process, which can bias veiling measurements where the IR excess is significant. For most sources, excess luminosity is only present in the infrared, outside of the BOSS spectral range. But in a handful of more extreme cases, the excess is prominent enough to be seen at the wavelength $<$ 1 micron which is roughly the wavelength limit of BOSS spectra. To address this, the SEDFit Python package was used \citep{sedfit}. This Python module queries VizieR to download all available photometric data for a source based on its coordinates and other stellar parameters such as \teff\ and \logg, and then performs SED fitting. The SED fitting process identifies the longest wavelength that is not affected by IR excess for each of the targeted sources by comparing the real data with a model. Thus, it is possible to determine the wavelength range from which the spectrum is affected by the IR excess. Veiling measurements were then limited to this unaffected wavelength range to ensure accuracy in the succeeding analysis.

Another important consideration are the temperature dependent systematics from the mismatch between the synthetic spectra and the real data of young stars. This is why we observed non-zero veiling in WTTSs and field stars, which do not have accretion disks and are not expected to exhibit veiling. For instance, the presence of spots in WTTSs can produce a multi-temperature spectrum that does not perfectly match the theoretical models, thereby introducing an excess veiling effect. 

Additional systematics are caused by imprecise measurements of continuum opacity, line opacity, and incomplete line list, which would affect the overall agreement between the synthetic template and the real observations of the flux.

To correct for all of these zero-point shifts in our veiling measurements, we subtracted the measured WTTS veiling (which is expected to be absent in them, and thus should primarily originate from the aforementioned systematics) from the CTTS veiling for each wavelength bins for all the sources. For this, we first divided the WTTSs veiling measurements in a set of Teff bins of 10 K, between 2300 and 11000 K, weighted them based on their uncertainties. Then, we subtracted the WTTSs veiling from the CTTSs veiling by matching the \teff\ of CTTSs to WTTSs. This correction ensures that the veiling measurements accurately reflect the properties of the CTTSs without being influenced by systematic errors present in the theoretical models (Figure \ref{fig:corrected_veiling}). 

\begin{figure}
\hspace{-0.3cm}
\centerline{
    \includegraphics[width=1\columnwidth,trim={0.24cm 0 0.2cm 0},clip]{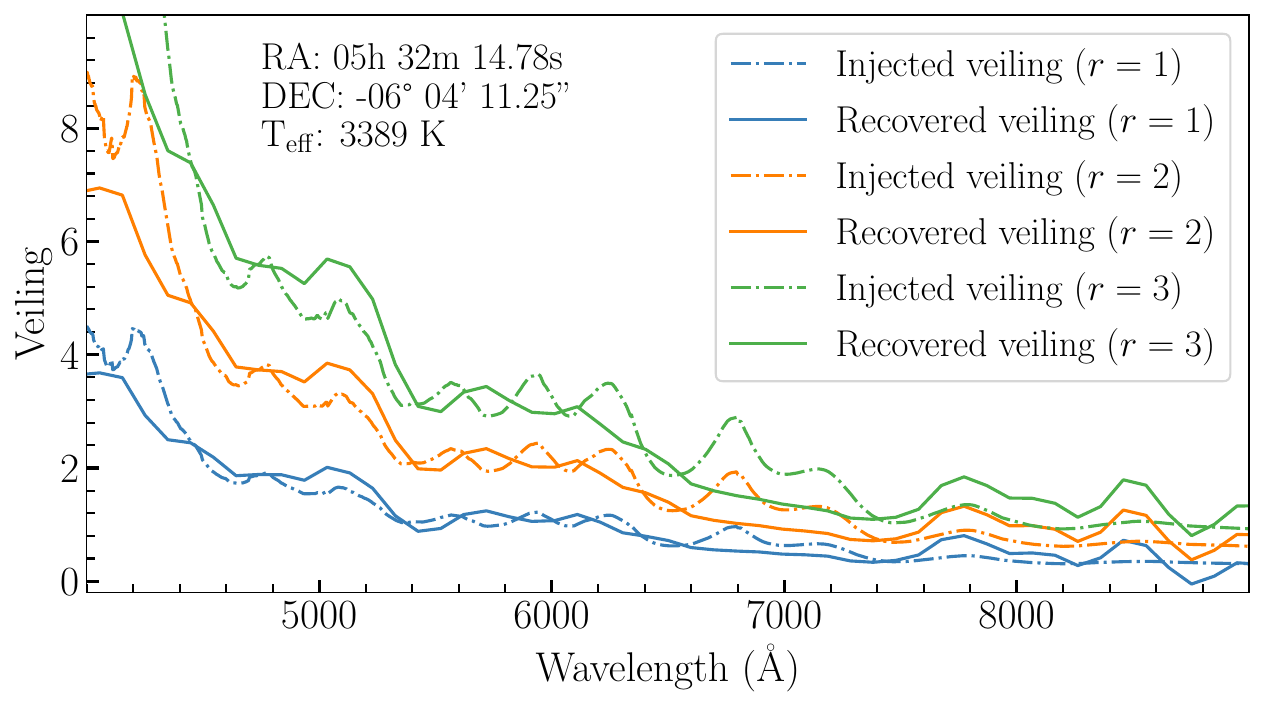} 
}
\caption{An example of of veiling recovery of artificially veiled WTTS spectra for one of the sources. We added constant excess continuum corresponding to veiling values of $r=1$, $2$, and $3$ relative to the median flux to representative WTTS spectra and processed them through our pipeline. Dashed lines show the originally injected veiling profiles and solid lines show the recovered veiling. The strong agreement supports the veracity of our methodology.
\label{fig:wtts_test}}
\end{figure}

To verify the effectiveness of this correction method, we applied artificial veiling to several WTTS through adding continuum to them. As the resulting veiling profile from that added continuum is known, we were able to compare it to the profile extracted with this pipeline, and we were able to recover the expected veiling. One of the sources that we selected for this test is shown in figure \ref{fig:wtts_test}.

\section{Results} \label{sec:results}

We expect to observe differences in the accretion properties in Classical T Tauri Stars (CTTSs) based on various factors such as effective temperature (\teff), the star's age, and the equivalent width of certain accretion sensitive lines like H$\alpha$ in the spectra. The veiling and its dependence on these factors can also vary across different wavelength ranges. We explore these aspects in detail.

\subsection{Comparing with the previous studies}

\begin{figure*}
\epsscale{1.2}
\plotone{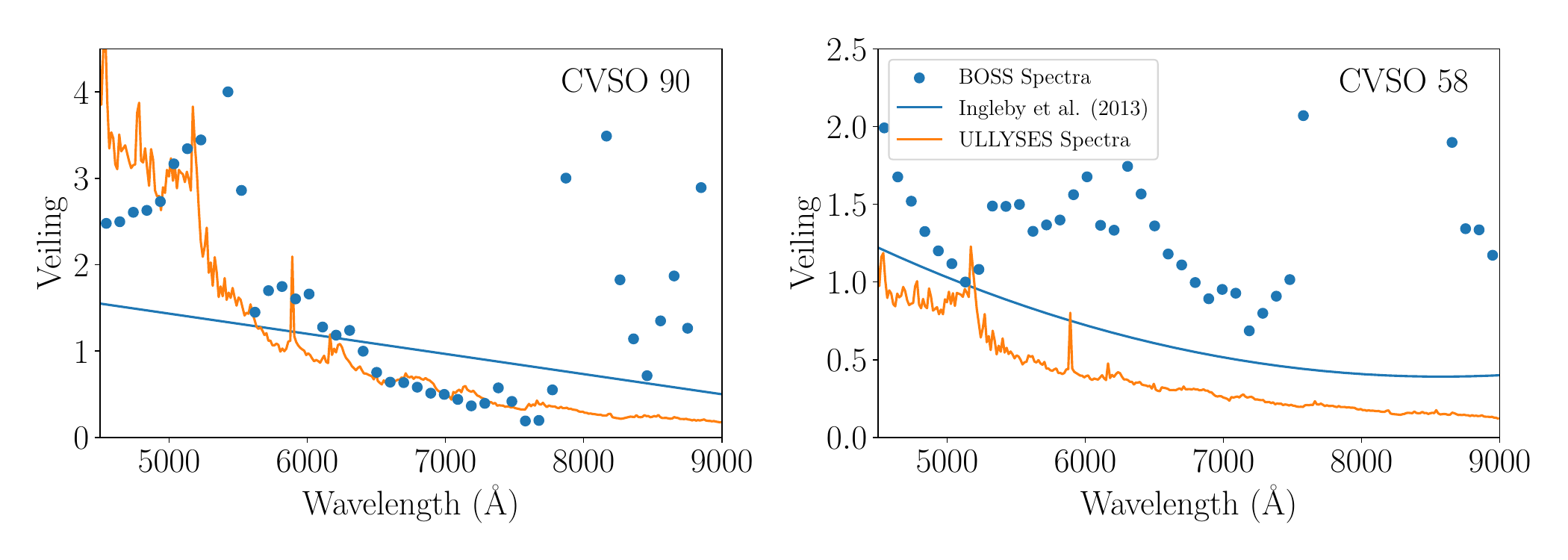}
\caption{Matched sources from \cite{ingleby2013}. The scatter plot represents the veiling data from our study, the orange line represents the veiling data from ULLYSES spectra in Pittman et al. (2025, in prep), and the blue line represents the recreation of the best-fit line from \cite{ingleby2013}.
\label{fig:matched_sources}}
\end{figure*}

A few previous studies have explored the relationship between veiling and wavelength. Among these, \cite{basri1990hamilton} and \cite{ingleby2013} stand out as significant contributions because of the resolution they achieved. To compare our findings, we searched for the sources studied in these works within our sample. While no matches were found with the sources from \cite{basri1990hamilton}, we successfully identified two sources, CVSO 90 and CVSO 58, from the study of \cite{ingleby2013}.

We attempted to approximate the veiling trends for these two sources as presented in \cite{ingleby2013} to compare them them with our measurements in Figure \ref{fig:matched_sources}. The curved line for CVSO 58 may not perfectly replicate the original data, but it is a good representation of their result. To ensure consistency, we constrained our analysis to the same wavelength ranges used in their study.

We used the following two equations to recreate the veiling trends:

For CVSO 90: 
\begin{equation} 
r_{\lambda} = -0.000233 \cdot \lambda + 2.6 
\end{equation}

For CVSO 58: 
\begin{align} 
r_{\lambda} = &-0.0001575 \cdot \lambda + 1.6175 \\
    &+ 5 \times 10^{-8} \cdot (\lambda - 7000)^2
\end{align}

\noindent Here, $\lambda$ is wavelength in \AA\ and $r_{\lambda}$ is the veiling.

Additionally, Pittman et al. (2025, in prep, private communication) have measured veiling for these two sources from ULLYSES spectra, obtained with the Hubble Space Telescope (HST). Their measurements are also included in Figure \ref{fig:matched_sources}.

We do find a broad agreement between the observations. For CVSO 58, although the normalization of the veiling is different between each work, the slope of the veiling profile as a function of wavelength has remained broadly similar. On the other hand, for CVSO 90, the veiling profile in this work matches the profile from Pittman et al both in magnitude and in shape, rapidly rising towards $r_\lambda\sim3$ towards the blue end, but \citet{ingleby2013} have reported a significantly flatter and more gradual $r_\lambda$.

We note that none of these observations were contemporaneous -- \citet{ingleby2013} have observed these sources in 2011. ULLYSES spectra were obtained in 2020. And BOSS spectra were obtained in 2021 for CVSO 58, and in 2023 for CVSO 90. Veiling is a highly variable quantity, driven by changes in accretion rates over time \citep{basri1990hamilton, calvet1998}, thus, it is reasonable to expect differences between these measurements. Significant variability may be expected even on very short time frame due to the observed geometry, whether an accretion hot spot is pointed towards or away from the observer. Given this, it is difficult to interpret which differences may be systematic in origin and which differences are physical.

In subsequent subsections we examine the systematic trends averaged out over a large sample of sources processed in a homogeneous manner, in an effort to minimize such temporal fluctuations.



\subsection{H$\alpha$ Equivalent Width}

Veiling is largely produced by accretion shocks where infalling material from the circumstellar disk impacts the stellar surface. This excess emission that originates from hot accretion columns and heated regions near the stellar surface can vary with temperature because the accretion process itself depends on spectral type.

The strength of H$\alpha$ line in a star depends on the accretion flow and the \teff\ of the system. The H$\alpha$ emission produced in the accretion columns and shocks results in intense accretion processes in CTTSs. Those CTTSs with strong H$\alpha$ emission thus will exhibit higher veiling. Also, H$\alpha$ emission can be related to the \teff\ of a star, so comparing the change of veiling with both H$\alpha$ eqw and \teff\ in different wavelength regions can be important.

\begin{figure}
\epsscale{1.2}
\plotone{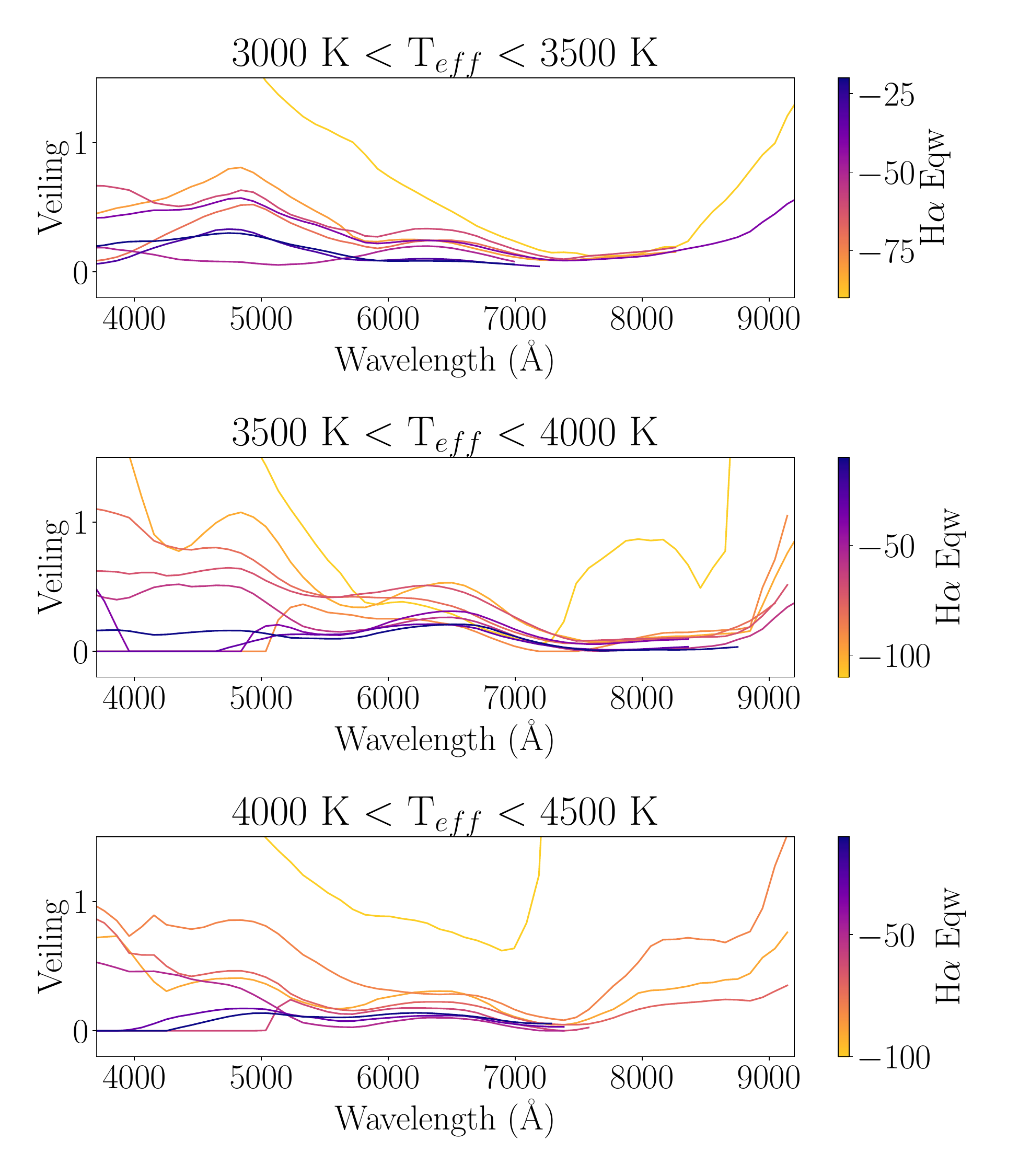}
\caption{Corrected Veiling of CTTSs color coded by the difference between their H$\alpha$ equivalent width and the CTTSs cutoff for different \teff\ ranges.
\label{fig:corrected_veiling_eqw}}
\end{figure}

To compare veiling across different temperatures, we calculated the average veiling of sources in discrete temperature bins. We separated the sources based on temperature ranges, with each bin covering a range of 100 K, considering temperatures from 3000 K to 5000 K. For each temperature bin, we calculated the average veiling array by taking the weighted mean veiling value at each wavelength across all sources within the bin, using the uncertainty in veiling as the weight. This provided a single average veiling array for each temperature range. Then we compared the veiling of CTTSs with the eqw of the H$\alpha$ line, by calculating the average veiling in bins defined by the equivalent width values. Each bin covered a range equal to 2 times the H$\alpha$ eqw. We then plotted the average veiling against wavelength for each eqw bin to visualize how veiling varies with both the width and the wavelength. Finally, we made three different plots for different \teff ranges starting from 3000 K to 4500 K each of them color coded by the H$\alpha$ eqw values. Figure \ref{fig:corrected_veiling_eqw} illustrates how veiling varies with both H$\alpha$ eqw and wavelength across three different \teff\ ranges. The veiling trends are consistent across all three temperature regions, showing three distinct peaks. The first peak occurs in the blue region, rising from the left of the plot and decreasing near 5000 \AA. The second peak is prominent in the mid-optical range, around 6000–7000 \AA, where veiling reaches to more higher values. Finally, there is a rise in veiling at the red end of the spectrum, beginning near 7500 \AA\ and continuing toward 9,000 \AA. When comparing veiling with H$\alpha$ eqw, the relationship becomes evident: sources with H$\alpha$ eqw values closer to 0 exhibit lower veiling, while sources with more negative H$\alpha$ eqw values show higher veiling across all three \teff\ ranges. This trend highlights the strong correlation between accretion activity, as indicated by H$\alpha$ emission, and veiling.

\subsection{Ages}

\begin{figure}
\epsscale{1.2}
\plotone{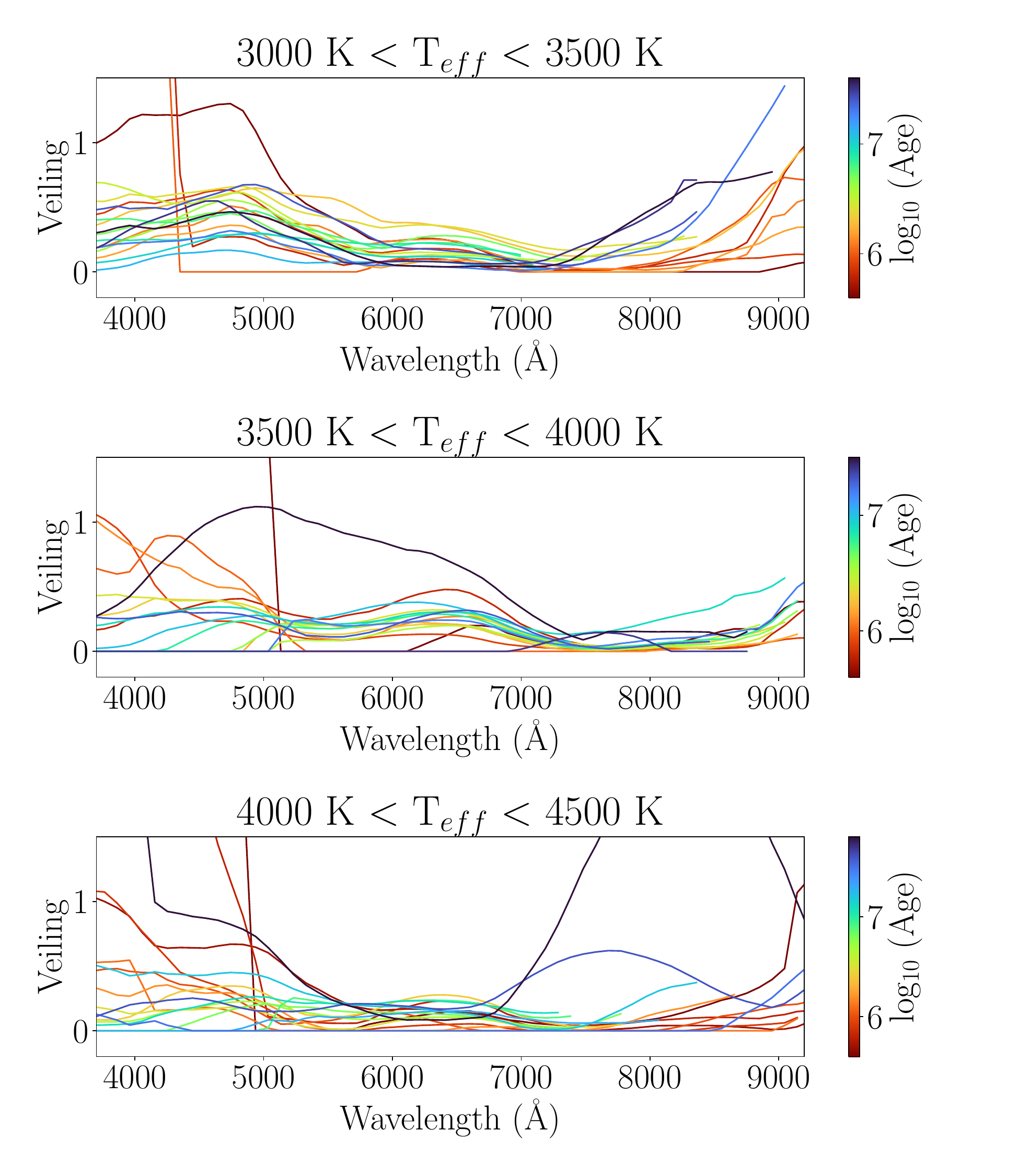}
\caption{Corrected Veiling of CTTSs color coded by age for different \teff\ ranges.
\label{fig:corrected_veiling_age}}
\end{figure}

Accretion flow can also vary with the age of CTTSs. During the pre-main sequence, where age varies from 0 to about 10 Myr, the accretion rate from their circumstellar disks can change. Typically, as the star evolves, the circumstellar disk also evolves, leading to a reduction in the mass accretion rate. This decline in accretion rate results in a reduction of the excess emission from accretion shocks, leading to lower veiling values in older stars. Conversely, if the accretion rate remains high or increases with age, the excess emission and veiling values can remain high or even increase.

To compare the veiling of CTTSs across different ages, we calculated the average veiling values in discrete age bins. Each age bin covered a range of 0.1 dex, and we considered ages from 1 to 30 Myr. For each age bin, we calculated the average veiling by taking the weighted mean veiling value at each wavelength for all sources within that bin, using the uncertainty in veiling as the weight. This provided a single average veiling array for each age range. We then plotted the average veiling against wavelength for each age bin in the previously defined \teff\ ranges. Figure \ref{fig:corrected_veiling_age} illustrates how veiling varies with both age and wavelength. Similar to figure \ref{fig:corrected_veiling_eqw}, the veiling trends with wavelength are consistent across all temperature ranges, showing three distinct regions of veiling increase: the blue region decreasing up to 5000 \AA; the mid-optical region, peaking between 6000–7000 \AA; and the red region, rising above 7500 \AA. When comparing veiling with age, however, no clear correlation emerges. The veiling values appear to vary randomly with age across the plotted temperature ranges.

\section{Discussion} \label{sec:discussion}

\subsection{Wavelength Dependence}

We empirically examined the dependence of the veiling profile on wavelength. A few previous studies have done similar examination on wavelength dependence of veiling, but most of them were done for only a handful of sources \citep{basri1990hamilton, Ingleby2014accretion, Muzerolle2003accretion}. Instead, we analyzed a comprehensive sample of 1800+ spectra, with varying \teff\ and evolution stages. As in previous studies, we found that veiling is not constant throughout the optical wavelength range. In fact, we found a number of features in the behavior. First it appears to increase sharply towards the UV, starting from 3700 \AA: our values peak near the blue limit of the BOSS spectra, likely increasing below that point. There is also a rise at the mid-optical range from 5000 \AA\ to 7000 \AA, peaking near 6000 \AA. Finally there is a rise from 8500 \AA\ to 10000 \AA, which peaks at the red limit of the BOSS spectra. These veiling peaks at different wavelength ranges evince several important physical processes.

The veiling rise at the UV end of the BOSS spectra from (3700 \AA\ to 5000 \AA) can be attributed to hot spots located at the accretion shock. These spots have a low filling factor, as the veiling increase in this part is much lower than the veiling increase in the mid-optical region of the spectra. However, due to their high temperature ($T$), even a small area ($A$) results in a substantial increase in luminosity ($L$), indicating that small, high-temperature regions can dominate the UV emission. Thus, these UV emitting hot spots likely have a small filling factor, covering a very limited area on the stellar surface. This suggests that these regions are likely where the accretion columns directly impact the stellar surface, creating localized, high-temperature spots that contribute to the observed veiling rise without needing a large emitting area.

In the mid-optical range (5000 \AA\ to 7000 \AA), the veiling increase is more pronounced compared to the veiling rise in other regions, indicating a higher filling factor. But in this range we observe a lower temperature contrast compared to the veling peak in the UV wavelength region, likely because the emitting spots in this range are cooler and closer to the stellar temperature. As luminosity ($L$) is proportional to surface area ($A$) and the fourth power of temperature ($T^4$), a lower temperature would require a significantly larger emitting area for the veiling effect to be noticeable. This means the veiling increase in this region is originated from a larger area on the surface of the star, which compensates for the lower temperature of the accretion related features. This increase can be attributed to hot spots located at the accretion shock. Such spots are created by the material accreting onto the star which results in excess emission. This excess emission can change the photospheric emission more than in other wavelength regions, hence the higher filling factor. The UV-emitting spots are possibly nested within the larger, cooler spots observed in the mid-optical range. The higher-energy spots appear to be embedded inside the broader spots of the mid-optical region.

At longer wavelengths, particularly in the 8500 \AA\ to 10,000 \AA\ range, the apparent increase in the veiling is likely not associated with accretion, but rather due to the presence of photospheric spots. Those spots would be different from other surface features, only contributing significantly to the veiling at these wavelengths. However, it is worth mentioning that our spectra present are significantly noisier at the longer wavelength regions, which may also cause a discrepancy in the veiling values at the longer wavelength range.

\subsection{H$\alpha$ Equivalent Width}

The H$\alpha$ eqw is an important spectral diagnostic for understanding the accretion processes in CTTSs. As the dominant hydrogen emission line, H$\alpha$ is a direct probe of the material flowing from the circumstellar disk onto the stellar surface. The strength of the H$\alpha$ emission is directly related to the amount of hydrogen present in the accretion stream. Therefore, stars with stronger accretion activity tend to show larger H$\alpha$ eqw values, which corresponds to higher mass accretion rates ($\dot{M}$). Numerous studies have demonstrated a clear relationship between H$\alpha$ emission and accretion, with stronger H$\alpha$ emission linked to higher accretion rates \citep{hartigan1994, white2003very, manara2014, wilson2022}.

In addition to its connection to accretion, the H$\alpha$ eqw provides insights into the temperature and density of the accretion flow. In the H$\alpha$ eqw plots (Figure \ref{fig:corrected_veiling_eqw}), we showed that stars with more negative eqw values exhibit higher veiling. This trend is more clear up to a wavelength of 7000 \AA\, and particularly in the wavelength regions where the veiling increases from the continuum. On one hand, we found that weaker H$\alpha$ sources tend to exhibit a weaker UV excess, with their veiling concentrated predominantly in the mid-optical wavelengths. On the other hand, the stars with stronger H$\alpha$ emission exhibit more pronounced veiling across the UV and optical ranges. In the \teff\ region of 3500 K to 4000 K of figure \ref{fig:corrected_veiling_eqw}, this trend is the most prominent. \cite{Ingleby2014accretion} found a similar trend where they compared energy flux of accretion shock model with normalized flux. All of this confirms that our results are consistent with expectations that stars with stronger H$\alpha$ emission, indicative of higher accretion rates, exhibit higher veiling values.

\subsection{Effective Temperature}

The effective temperature (\teff) of a star is also related to the H$\alpha$ emission. Although H$\alpha$ emission is commonly used as an indicator of accretion, \teff\ itself is not a direct indicator of the presence of an accretion disk.

In our study, we compared the veiling profiles across three different temperature ranges (3000 K $<$ \teff\ $<$ 3500 K, 3500 K $<$ \teff\ $<$ 4000 K and 4000 K $<$ \teff\ $<$ 4500 K) in both figure \ref{fig:corrected_veiling_eqw}, and figure \ref{fig:corrected_veiling_age}). In them we showed that the veiling shows no clear dependence on \teff\ as the veiling profile remains almost similar across all the three different \teff\ regions.

\subsection{Age}
Veiling can also have a relationship with the age of the CTTSs. This variation can be driven by various factors that influence the accretion processes throughout a CTTSs different stages of development. In theory, as CTTSs evolve and approach the later stages of their pre-main sequence phase, we would expect a gradual decrease in accretion rates as circumstellar disks dissipate \citep{strom1993, gullbring1998}. The circumstellar disk becomes more thin and decreases in density as the pre-main sequence star evolves. As a result, the star becomes less capable of sustaining high accretion rates. So, the older CTTSs should exhibit a lower veiling compared to the younger CTTSs.

However, the result from our analysis in figure \ref{fig:corrected_veiling_age} suggests that the relationship between veiling and age is more complex than a simple linear decrease. While it is true that, on average, older CTTSs tend to show lower veiling, there are notable exceptions. One such example can be the TW Hydra \citep{Herczeg2023} which is an old star but shows signicant accretion activity. Similarly, some younger CTTSs can also exhibit low accretion activity. This demonstrates that the dissipation of the disk and the decrease in veiling is not strictly age dependent. Rather, the accretion activity is subject to diverse factors. The scatter in accretion strengths across different ages is significant, even though the general trend points to a gradual decrease in accretion over time. This variability highlights that stars lose their disks in a probabilistic manner, with some stars retaining their disks and continuing to accrete for much longer periods, while others lose their disks rapidly within less than 1 Myr. Thus, while age correlates with a decrease in veiling, this trend is less pronounced than the more direct relationship between H$\alpha$ emission and accretion strength.

\section{Conclusions} \label{sec:conclusions}
We selected a few thousand BOSS spectra of pre-main sequence stars in order to understand their accretion activity through veiling. The source sample include CTTSs, WTTSs, as well as a control sub-sample of field stars. We calculated the extinction and de-reddened the spectra for all the sources. Then we measured veiling for a selected number of BOSS sources in discrete wavelength ranges. 

Next, we compared the veiling of CTTSs, WTTSs, and field stars. We found not null veiling for WTTSs and field stars, due to some systematics in our data. To correct the veiling of CTTSs for those systematics, we binned the veiling of WTTSs for discrete \teff\ bins and then subtracted the average WTTSs veiling from CTTSs. Then we compared the corrected veiling of CTTSs with different physical properties of the CTTSs. These properties include wavelength, H$\alpha$ eqw, \teff, and age of the stars.

We found that veiling presents a number of features in the BOSS wavelength range. Some of these features are attributed to the behavior of the accretion stream and spots on the accretion disk while some of these features are not related to the accretion activity of the stars. 

We also found veiling to be related to both H$\alpha$ eqw and \teff. More H$\alpha$ eqw means more veiling. On the other hand, some of the features in veiling were found to be more prominent in some \teff\ ranges. We also found that the peak of the veiling in mid-optical region shifts towards redder wavelengths for lower eqw values. 

Finally, we compared veiling with stellar age, and found that veiling varies as the star evolves. Sometimes some old stars can still remain active and show high veiling while some stars can show low veiling even if they are very young.

\software{TOPCAT \citep{topcat}, BOSS Net \citep{bossnet}}, AstroPy \citep{astropy}, LineForest \citep{lineforest}, Sagitta \citep{McBride2021sagitta}

\acknowledgments

We thank Caeley Pittman and Nuria Calvet for their input on the manuscript.
We thank Sean Morrison for the input regarding the flux calibration of BOSS spectra.

MK acknowledges support provided by NASA grant 80NSSC24K0620. CRZ acknowledge support from project UNAM-DGAPA-PAPIIT IN101723. R. L-V acknowledges support from Secretar\'ia de Ciencia, Humanidades, Tecnolog\'ia e Inovacci\'on (SECIHTI) through a postdoctoral fellowship within the program ``Estancias posdoctorales por M\'exico''

Funding for the Sloan Digital Sky Survey V has been provided by the Alfred P. Sloan Foundation, the Heising-Simons Foundation, the National Science Foundation, and the Participating Institutions. SDSS acknowledges support and resources from the Center for High-Performance Computing at the University of Utah. SDSS telescopes are located at Apache Point Observatory, funded by the Astrophysical Research Consortium and operated by New Mexico State University, and at Las Campanas Observatory, operated by the Carnegie Institution for Science. The SDSS web site is \url{www.sdss.org}.

SDSS is managed by the Astrophysical Research Consortium for the Participating Institutions of the SDSS Collaboration, including the Carnegie Institution for Science, Chilean National Time Allocation Committee (CNTAC) ratified researchers, Caltech, the Gotham Participation Group, Harvard University, Heidelberg University, The Flatiron Institute, The Johns Hopkins University, L'Ecole polytechnique f\'{e}d\'{e}rale de Lausanne (EPFL), Leibniz-Institut f\"{u}r Astrophysik Potsdam (AIP), Max-Planck-Institut f\"{u}r Astronomie (MPIA Heidelberg), Max-Planck-Institut f\"{u}r Extraterrestrische Physik (MPE), Nanjing University, National Astronomical Observatories of China (NAOC), New Mexico State University, The Ohio State University, Pennsylvania State University, Smithsonian Astrophysical Observatory, Space Telescope Science Institute (STScI), the Stellar Astrophysics Participation Group, Universidad Nacional Aut\'{o}noma de M\'{e}xico, University of Arizona, University of Colorado Boulder, University of Illinois at Urbana-Champaign, University of Toronto, University of Utah, University of Virginia, Yale University, and Yunnan University.

This work has made use of data from the European Space Agency (ESA)
mission {\it Gaia} (\url{https://www.cosmos.esa.int/gaia}), processed by
the {\it Gaia} Data Processing and Analysis Consortium (DPAC,
\url{https://www.cosmos.esa.int/web/gaia/dpac/consortium}). Funding
for the DPAC has been provided by national institutions, in particular
the institutions participating in the {\it Gaia} Multilateral Agreement.

\bibliographystyle{aasjournal.bst}
\bibliography{references.bib}

\end{document}